\documentstyle[preprint,aps]{revtex}
\newcommand{\be}{\begin{equation}} \newcommand{\ee}{\end{equation}} 
\newcommand{\bea}{\begin{eqnarray}}\newcommand{\eea}{\end{eqnarray}}

\begin{document}
\draft
\preprint{MRI-PHY/96-09, hep-th/9603122}
\title { Self-dual Gauged $CP^N$ Models}
\author{Pijush K. Ghosh$^{*}$}
\address{The Mehta Research Institute of
Mathematics \& Mathematical Physics,\\
Allahabad-211002, INDIA.}
\footnotetext {$\mbox{}^*$ E-mail: 
pijush@mri.ernet.in }  
\maketitle
\begin{abstract} 
We consider a $CP^N$ model with the subgroup $SU(r)$ completely gauged, 
where $1 < r < N+1$. The gauge field dynamics is solely governed by a
nonabelian Chern-Simons term and the global $SU(N+1)$ symmetry is broken
explicitly by introducing a $SU(r)$ invariant scalar potential. We obtain
self-dual equations of this gauged $CP^N$ model and find that the energy 
is bounded from below by a linear combination of the topological charge
and a global $U(1)$ charge present in the theory. We also discuss on
the self-dual soliton solutions of this model.
\end{abstract}
\narrowtext

\newpage

The self-dual soliton solutions of the $CP^1$ model
\cite{sig} in $2+1$ dimensions are scale invariant.
Due to this conformal invariance, the size of these solitons can change
arbitrarily during the time evolution without costing any energy.
Naturally, the particle interpretation of these solitons upon quantization
is not valid. There are several ways to break the scale invariance of this
model. One such remedy is to make the global symmetry local by introducing
a dynamical gauge field and to incorporate scalar potential in the theory.
Gauging of the global symmetry group is a natural one also as the fundamental
interactions in nature are mediated by gauge bosons. However, no potential
term can be introduced
maintaining the $SU(2)$ invariance since the complex doublet has unit
norm in the internal space. The self-dual soliton solutions of completely
gauged $CP^1$ models \cite{nardeli}, even though known exactly, are also
scale invariant.

Naturally, one is led to consider $CP^1$ models with a dynamical $U(1)$
gauge field where a $U(1)$ invariant potential can be introduced
\cite{may,dur,gg,itf}.
There are two different approaches in gauging the $CP^1$
model. One is to replace the auxiliary $U(1)$ gauge field by a
dynamical one \cite{may}. However, no model based on such a gauging
prescription is known to admit self-dual soliton solutions. On the other
hand, self-dual solution is possible, if a $U(1)$ subgroup of the
global $SU(2)$ symmetry is gauged, leaving the auxiliary $U(1)$ gauge
field unchanged \cite{dur,gg,itf}\footnote{Both the references \cite{dur}
and \cite{gg} describe $U(1)$ gauged $O(3)$ sigma models in $2+1$
dimensions. Here, we are mentioning the equivalent $CP^1$ version
of these models.}. Such self-dual gauged $CP^1$ models
offer a variety of new soliton solutions. For example, the energy of
the self-dual soliton
solutions are infinitely degenerate in each topological sector. This is a
manifestation of the fact that the magnetic flux is not necessarily quantized
in terms of the topological charge, even though the energy is quantized
\cite{dur,gg}.
Also, such gauged $CP^1$ models admit both topological as well as
nontopological self-dual soliton solutions for a specific choice of the
potential, when the gauge field dynamics is solely governed by a Chern-Simons
(CS) term \cite{gg}. These static solitons acquire nonzero charge and angular
momentum due to the presence of the CS term. Both the charge as
well as the angular momentum of these soliton solutions are fractional
in general and thereby a good candidate for anyons.

The purpose of this letter is to generalize the results of self-dual gauged
$CP^1$ models to the $CP^N$ case for arbitrary $N$.
In particular, we consider a $CP^N$ model with the subgroup $SU(r)$
completely gauged, where $1 < r < N+1$.
The gauge field dynamics is solely governed by a nonabelian
CS term and the complex scalar fields interact among themselves
through a $SU(r)$ invariant scalar potential. We obtain self-dual equations
of this model. Unlike the case of gauged $CP^1$ model, we find that
the energy is bounded from below by a linear combination of
the topological charge and a global $U(1)$ charge present in the
theory. We also discuss qualitatively about the solitons of this model.

The $CP^N$ Lagrangian is given by,
\be
{\cal{L}} = {\mid \partial_\mu \eta \mid}^2 - {\mid \eta^\dagger
\partial_\mu \eta \mid}^2,
\label{eq0}
\ee
\noindent where $\eta$ is a $(N+1)$-component complex scalar field with
the constraint $ \eta^{\dagger} \eta =1$. The Lagrangian (\ref{eq0}) has
a global $SU(N+1)$ as well as a local $U(1)$ invariance. The static,
self-dual equation of this model is,
\be
( {\cal{D}}_i \pm i \epsilon_{ij} {\cal{D}}_j ) \eta = 0,
\label{eq0.0}
\ee
\noindent where $ {\cal{D}}_\mu \eta = \partial_\mu \eta - (\eta^\dagger
\partial_\mu \eta ) \eta$. Note that $\eta^\dagger \partial_\mu \eta$
can be interpreted as an auxiliary $U(1)$ gauge field. The most general
solution for $\eta$ can be
written down in terms of $N$ (anti)holomorphic functions $w_\alpha$,
\be
\eta = \frac{1}{(1 + {\mid w \mid}^2)^{\frac{1}{2}}}
\pmatrix{ {w} \cr {1}},
\label{eq0.1}
\ee
\noindent where $w$ is a vector with $N$ components, $w_\alpha$.
These solitons are characterized by the energy $E= 2 \pi {\mid T \mid}$
where the topological charge is determined as $T=\int d^2x k_0$.
Here, $k_0$ is the zeroth component of the topological
current $k_\mu$,
\be
k_\mu = \frac{1}{2 \pi i} \epsilon_{\mu \nu \lambda} 
( {\cal{D}}^\nu \eta )^\dagger ( {\cal{D}}^\lambda \eta ).
\label{eq1}
\ee
\noindent Note that the topological current is conserved.

We are interested in gauging a subgroup $SU(r)$ of the global $SU(N+1)$
group of the $CP^N$ model (\ref{eq0}), where $1 < r < N+1$. For this
purpose, let us decompose $\eta$ as
$ \eta^\dagger = ( \psi^\dagger, \phi^\dagger )$, where $\psi$ and $\phi$
have $r$ and $q$ components respectively such that
$r + q = N+1$. The covariant derivative is defined as,
\be
D_\mu \eta = \pmatrix{{D_\mu \psi} \cr {\partial_\mu \phi}}
= \pmatrix{{(\partial_\mu - i A_\mu^a \Gamma^a) \psi} \cr 
{\partial_\mu \phi}} ,
\label{eq2}
\ee
\noindent where $A_\mu^a$ are the $SU(r)$ gauge fields.
The $SU(r)$ generators
$\Gamma^a$'s are hermitian traceless matrices with the properties, (i)
$[\Gamma^a,\Gamma^b] = i f^{abc} \Gamma^c$ and (ii) $Tr(\Gamma^a \Gamma^b)
=\frac{1}{2} \delta^{ab}$. 
Now replacing $\partial_\mu \eta$ by $D_\mu \eta$ in
(\ref{eq0}) and
introducing a CS term for the gauge field dynamics as well as a $SU(r)$
invariant scalar potential, we have the following
Lagrangian, 
\bea
{\cal{L}} \ & = & \ {\mid D_\mu \eta  \mid}^2 -
{\mid \eta^\dagger D_\mu \eta \mid}^2 
+ \frac{\kappa}{4} \epsilon^{\mu \nu \lambda} \left
( F_{\mu \nu}^a A_\lambda^a - \frac{1}{3}
f^{a b c} A_\mu^a A_\nu^b A_\lambda^c \right )\nonumber \\
& &  -  \beta^2 {\mid \psi \mid}^2 {\mid \phi \mid}^2 \left (
{\mid \psi \mid}^2 - v^2 \right )^2 ,
\label{eq3.0}
\eea
\noindent where $\beta=\frac{r-1}{2 r \kappa}$. We work here in
Minkowskian space-time with the signature
$g_{\mu \nu}=( 1, -1, -1 )$ and the field-strength $F_{\mu \nu}^a$ is
$F_{\mu \nu}^a = \partial_\mu A_\nu^a - \partial_\nu A_\mu^a + 
f^{a b c} A_\mu^b A_\nu^c $. The velocity of light $c$ and the
Planck's constant in units of $\frac{1}{2 \pi}$ are taken to be unity.
The coefficient of the CS term ($\kappa$) has inverse mass dimension.
 
The particular form of the scalar potential in (\ref{eq3.0}) is completely
determined by the self-duality. The global $SU(N+1)$ invariance of the
$CP^N$ model is lost because of the introduction of this scalar potential.
However, the local $U(1)$ invariance
of the Lagrangian (\ref{eq0}) under the transformation
$\eta \rightarrow \eta e^{i \alpha(x)}$ is still
maintained. Exploiting this $U(1)$ invariance,
the Lagrangian (\ref{eq3.0}) can be conveniently rewritten as,
\be
{\cal{L}} = {\mid \bigtriangledown_\mu \eta  \mid}^2 
+ \frac{\kappa}{4} \epsilon^{\mu \nu \lambda} \left
( F_{\mu \nu}^a A_\lambda^a - \frac{1}{3}
f^{a b c} A_\mu^a A_\nu^b A_\lambda^c \right )
- \beta^2 {\mid \psi \mid}^2 {\mid \phi \mid}^2 \left (
{\mid \psi \mid}^2 - v^2 \right )^2,
\label{eq3}
\ee
\noindent where we define $\bigtriangledown_\mu \eta = D_\mu \eta
-(\eta^\dagger D_\mu \eta) \eta $  as in the case of the usual
$CP^N$ model. Note that the Lagrangian (\ref{eq3}) reduces to the
usual $CP^N$ model (\ref{eq0}) in the limit $\kappa \rightarrow \infty$,
i. e. $\beta \rightarrow 0$. In this limit, the potential term drops
out from the action and the equations of motion for the gauge fields
dictate that they are pure gauges, $F_{\mu \nu}^a = 0$. As a result,
we are left with the usual $CP^N$ model.

The scalar potential has three degenerate minima for $0 < v^2 < 1$,
namely, (i) $\psi=0$, (ii) ${\mid \psi \mid}= 1$ and (iii)
${\mid \psi \mid}= v$. Note that ${\mid \psi \mid}= 1$ is a 
minimum because of the constraint $\eta^{\dagger} \eta = 1$. These three
degenerate minima reduces to two in case $v^2 \geq 1$ or $v^2 = 0$.
$\psi=0$ corresponds to the
symmetric phase of the theory. However, the $SU(r)$ symmetry is
spontaneously broken
to $SU(r-1)$, when ${\mid \psi \mid}= v$ with the
restriction $0 < v^2 < 1$  or ${\mid \psi \mid}=1$.
In the asymmetric phase 
${\mid \psi \mid} = v$, there are (i) $r^2 - 2 r$ massless gauge bosons,
(ii) $2 r - 1$ massive gauge bosons and (iii) one massive Higgs particle.
The particle content of the asymmetric phase ${\mid
\psi \mid} = 1$ is different from this. In particular, we have (i)
$(r-1)^2$ massless gauge bosons, (ii) $2 r - 2$ massive gauge bosons
and (iii) one massive Higgs particle. In this phase, no degree
of freedom is left for the gauge field associated with the broken
diagonal generator $\Gamma^D$ to become massive.

The gauge-invariant topological current for the model (\ref{eq3})
can be defined as,
\be
K_\mu = \frac{1}{2 \pi i} \epsilon_{\mu \nu \lambda} 
\left [ (\bigtriangledown^\nu \eta )^\dagger
(\bigtriangledown^\lambda \eta ) - \frac{i}{2} 
F^{a,\nu \lambda} \psi^\dagger \Gamma^a \psi \right ] .
\label{eq3.1}
\ee
\noindent The last term in (\ref{eq3.1}) is introduced to make
the topological current $K_\mu$ divergence free. In particular, $K_\mu$
differs from the topological current $k_\mu$ of the usual $CP^N$ model
by a total divergence,
\be
K_\mu = k_\mu - \frac{1}{2 \pi} \epsilon_{\mu \nu \lambda}
\partial^\nu \left
( A^{a,\lambda} \psi^\dagger \Gamma^a \psi \right ).
\label{eq3.2}
\ee
\noindent The nontrivial mapping $\Pi_2(CP^N) = Z$ is characterized
by the first term of Eq. (\ref{eq3.2}).
We now define a gauge-invariant abelian field-strength
${\cal{F}}_{\mu \nu}$ in terms of the topological current $K_\mu$,
\be
{\cal{F}}_\mu = \frac{1}{2} \epsilon_{\mu \nu \lambda} {\cal{F}}^
{\nu \lambda} = K_\mu .
\label{eq3.3}
\ee
\noindent Note that the topological charge $Q=\int d^2 x K_0$ is
exactly equal to the magnetic flux $\Phi=\int d^2 x {\cal{F}}_{12}$.
Also, the conservation of the topological current automatically
implies the Bianchi identity $\partial_\mu {\cal{F}}^\mu = 0$. 

The equations of motion which follow from the Lagrangian (\ref{eq3})
are,
\bea
\bigtriangledown_\mu \bigtriangledown^\mu \psi
\ & = & \ \left ( \eta^\dagger \bigtriangledown_\mu \bigtriangledown^\mu
\eta \right ) \psi\nonumber \\
& & + \beta^2
{\mid \phi \mid}^2 \left ( {\mid \psi \mid}^2 - v^2 \right )  \bigg (
2 {\mid \psi \mid}^2
\left ( 2 {\mid \psi \mid}^2 - v^2 \right ) - \left ( 3 {\mid \psi \mid}^2 -
v^2 \right )  \bigg ) \psi,
\label{eq4}
\eea
\bea
\bigtriangledown_\mu \bigtriangledown^\mu \phi
\ & = & \ \left ( \eta^\dagger \bigtriangledown_\mu \bigtriangledown^\mu
\eta \right ) \phi\nonumber \\
& & + \beta^2
{\mid \psi \mid}^2 \left ( {\mid \psi \mid}^2 - v^2 \right )  \bigg ( 2
{\mid \phi \mid}^2
\left ( 2 {\mid \psi \mid}^2 - v^2 \right ) - \left ( {\mid \psi \mid}^2 - 
v^2 \right )  \bigg ) \phi,
\label{eq5}
\eea
\be
\frac{\kappa}{2} \epsilon_{\mu \nu \lambda} F^{a, \nu \lambda} =
J_\mu^a = - i \bigg ( \psi^\dagger \Gamma^a  \left ( \bigtriangledown_\mu
\psi 
\right ) - \left ( \bigtriangledown_\mu \psi \right )^\dagger \Gamma^a \psi
\bigg ),
\label{eq6}
\ee
\noindent where $J_\mu^a$ is the Noether current associated with the
$SU(r)$ invariance. The constraint $\eta^\dagger \eta=1$ has been taken
into considerations in deriving Eqs. (\ref{eq4}) and (\ref{eq5}), though
not mentioned explicitly in (\ref{eq3}). Multiplying the Gauss law by
$\psi^\dagger \Gamma^a \psi$ and using the identity,
\be
\sum \left ( \Gamma^a \right )_{ij} \left ( \Gamma^a \right )_{k l}
= \frac{1}{2} \delta_{i l} \delta_{j k} - \frac{1}{2 r}
\delta_{ij} \delta_{kl},
\label{eq7}
\ee
\noindent we have,
\be
F_{12}^a \psi^\dagger \Gamma^a \psi = \beta  j_0  {\mid \psi \mid}^2,
\ \ j_0 = - i \left [ \psi^\dagger (\bigtriangledown_0 \psi )
- (\bigtriangledown_0 \psi )^\dagger \psi \right ],
\label{eq8}
\ee
\noindent where $j_0$ is the global $U(1)$ current associated with the
transformation $ \psi \rightarrow e^{i \gamma} \psi $.

The energy functional $E$ can be obtained by varying (\ref{eq3}) with
respect to the background metric,
\be
E = \int d^2 x \left [ {\mid \bigtriangledown_0 \eta \mid }^2
+ {\mid \bigtriangledown_i \eta \mid }^2 + 
\beta^2 {\mid \psi \mid}^2 {\mid \phi \mid}^2 \left (
{\mid \psi \mid}^2 - v^2 \right )^2 \right ].
\label{eq9}
\ee
\noindent Note that the CS term, being a topological term, has
no contribution to the energy functional. Using the identities, 
\be
{\mid \bigtriangledown_i \eta \mid }^2 = {\mid \left ( \bigtriangledown_1
\pm i \bigtriangledown_2 \right ) \eta \mid }^2 \mp i \epsilon_{jk}
(\bigtriangledown_j \eta)^\dagger (\bigtriangledown_k \eta),
\label{eq10}
\ee
\be
\left ( \bigtriangledown_\mu \psi \right )^\dagger \psi
= - \left ( \bigtriangledown_\mu \phi \right )^\dagger \phi
= {\mid \phi \mid}^2 \left ( D_\mu \psi \right )^\dagger \psi
- {\mid \psi \mid}^2 \partial_\mu \phi^\dagger \phi, 
\label{eq10.0}
\ee
\bea
{\mid \bigtriangledown_0 \eta \mid }^2 + 
\beta^2 {\mid \psi \mid}^2 {\mid \phi \mid}^2 \left (
{\mid \psi \mid}^2 - v^2 \right )^2 \ & = & \ 
{\mid \left ( \bigtriangledown_0 \pm i  
\beta \left ( {\mid \psi \mid}^2 - v^2 \right )
{\mid \phi \mid}^2 \right ) \psi \mid}^2\nonumber \\
& & + {\mid \left ( \bigtriangledown_0 \mp i  
\beta \left ( {\mid \psi \mid}^2 - v^2 \right )
{\mid \psi \mid}^2 \right ) \phi \mid}^2\nonumber \\
& & \mp \beta  j_0 \left ( {\mid \psi \mid}^2 - v^2 \right ),
\label{eq10.1}
\eea
\noindent the energy functional (\ref{eq9}) can be rewritten as, 
\bea
E & \ = & \ \int d^2 x \bigg [ {\mid \left ( \bigtriangledown_1 \pm 
i \bigtriangledown_2
\right ) \eta \mid }^2 + {\mid \left ( \bigtriangledown_0 \pm i  
\beta \left ( {\mid \psi \mid}^2 - v^2 \right )
{\mid \phi \mid}^2 \right ) \psi \mid}^2\nonumber \\
& & +  {\mid \left ( \bigtriangledown_0 \mp i  
\beta \left ( {\mid \psi \mid}^2 - v^2 \right )
{\mid \psi \mid}^2 \right ) \phi \mid}^2 \bigg ] \pm 2 \pi Q
\pm \beta v^2 {\tilde{Q}},
\label{eq11}
\eea
\noindent where $Q=\int d^2 x K_0$ is the topological charge and
${\tilde{Q}}= \int d^2 x j_0$ is the global $U(1)$ charge. Note
that the relative sign between the last two terms in (\ref{eq11})
is uniquely fixed by the self-duality.
The energy in Eq. (\ref{eq11}) has a lower bound
$E \geq {\mid 2 \pi Q + \beta v^2 {\tilde{Q}} \mid}$ in terms of the
topological and the global $U(1)$ charge. This is unlike the case
of gauged $CP^1$ model \cite{gg} where this Bogomol'nyi bound is given
in terms of the topological charge only. It is also
worth recalling at this point that the energy of the self-dual
non-abelian CS solitons \cite{klee} is bounded from below by a global
$U(1)$ charge only. Thus, in contrast to the self-dual nonabelian CS
case, we expect our model to admit a variety of soliton solutions.

The bound $E \geq {\mid 2 \pi Q + \beta v^2 {\tilde{Q}} \mid}$ is
saturated when the following Bogomol'nyi equations \cite{bogo}
are satisfied\footnote{
The self-dual equations for the $U(1)$ gauged $CP^1$ model of Ref.
\cite{gg} can be obtained from these Bogomol'nyi equations
by replacing $(D_\mu \eta)^T  \rightarrow ( (\partial_\mu - i a_\mu) \psi,
\partial_\mu \phi)$, $A_\mu^a \rightarrow a_\mu$, $\Gamma^a \rightarrow 1$,
$\beta \rightarrow \frac{1}{\kappa}$ and fixing $r, q = 1$.},
\be
\left ( \bigtriangledown_1 \pm 
i \bigtriangledown_2 \right ) \eta = 0 ,
\label{eq12}
\ee
\be
\bigg ( \bigtriangledown_0 \pm i  
\beta ( {\mid \psi \mid}^2 - v^2 )
{\mid \phi \mid}^2 \bigg ) \psi = 0 ,
\label{eq13}
\ee
\be
\bigg ( \bigtriangledown_0 \mp i  
\beta ( {\mid \psi \mid}^2 - v^2 )
{\mid \psi \mid}^2 \bigg ) \phi = 0 .
\label{eq14}
\ee
\noindent In order to ensure that these first order equations are
consistent
with the second order field equations, first note that for static
field configurations, both Eqs. (\ref{eq13}) as well as (\ref{eq14})
imply,
\be
A_0^a \psi^\dagger \Gamma^a \psi = \pm \beta \left ( {\mid \psi \mid}^2
- v^2 \right ) {\mid \psi \mid}^2 .
\label{eq15}
\ee
\noindent The above equation can be solved as,
\be
A_0^a = \pm \frac{1}{\kappa} \frac{\psi^\dagger \Gamma^a \psi}{
{\mid \psi \mid}^2} \left ( {\mid \psi \mid}^2 - v^2 \right ) .
\label{eq16}
\ee
\noindent On the other hand, using Eqs. (\ref{eq6}), (\ref{eq7}) and
(\ref{eq13}),
$F_{12}^a$ is completely determined in terms of the complex scalar
fields,
\be
F_{12}^a = \mp \frac{2 \beta}{\kappa} \left ( {\mid \psi \mid}^2
- v^2 \right )  {\mid \phi \mid}^2 
\psi^\dagger \Gamma^a \psi .
\label{eq17}
\ee
\noindent With the help of Eqs. (\ref{eq16}), (\ref{eq17}) and the
identity $\epsilon_{ij} D_i D_j \psi = - i F_{12}^a \Gamma^a \psi$,
it can be shown after a lengthy calculation that Eqs. (\ref{eq4})
and (\ref{eq5}) are indeed consistent with the first order self-dual
equations.
 
Let us now discuss about the nature of these self-dual soliton solutions.
The energy is determined solely in terms of the topological charge
$Q$, when
$v^2=0$. The topological charge can receive contribution either from
the first or the last term in (\ref{eq3.2}). 
In the former case, the stability of these solitons are guaranteed by
topological arguments. This is not true for the later situation and
thus, are nontopological solitons.
When $v^2 \neq 0$, the self-dual solitons can be either of Q-lumps
\cite{rl} or Q-balls \cite{sc} type, depending on whether $Q$
receives contribution from the first or the last term of (\ref{eq3.2})
respectively. These self-dual solitons are of Q-lumps type, when
$Q$ receives contribution from both the terms of (\ref{eq3.2})
and $v^2 \geq 0$.
 
In order to analyze the self-dual equations, we introduce two
new variables,
\be
W_\psi = \frac{\psi}{\phi_q} , \ \ 
W_\phi = \frac{\phi}{\phi_q} ,
\label{eq19}
\ee
\noindent where $\phi_q$ is the $q$th component of $\phi$. Both $W_\psi$
and $W_\phi$ are invariant under the simultaneous transformations
$ \psi \rightarrow e^{i \alpha(x)} \psi$ and $\phi \rightarrow
e^{i \alpha(x)} \phi$. Note that the $q$th component of $W_\phi$ is 
equal to unity and ${\mid \phi_q
\mid }^2 {\mid W \mid}^2 = 1$ where $ {\mid W \mid}^2 =
{\mid W_\psi \mid}^2 + {\mid W_\phi \mid}^2$. With the help
of this, the self-dual equations can be conveniently written as,
\bea
& & \left ( D_1 \pm i D_2 \right )
W_\psi=0, \ \ \left ( \partial_1 \pm i \partial_2 \right )
W_\phi = 0,\nonumber \\
& & F_{12}^a = \mp \frac{2 \beta}{\kappa} \frac{{\mid W_\phi \mid}^2}
{{\mid W \mid}^6} \left ( {\mid W_\psi \mid}^2 - v^2 {\mid W \mid}^2
\right ) W_\psi^\dagger \Gamma^a W_\psi.
\label{eq20}
\eea
\noindent The second equation of (\ref{eq20}) implies that
all the components of $W_\phi$ are arbitrary (anti)holomorphic
functions. On the other hand, the gauge potentials can be determined
from the first Eq. of (\ref{eq20}) as,
\be
A_\pm^a = - \frac{i}{\kappa \beta} \frac{1}{{\mid W_\psi \mid}^2}
W_\psi^\dagger \Gamma^a \partial_\pm W_\psi, 
\label{eq21}
\ee
\noindent where $A_\pm^a = A_1^a \pm i A_2^a$ and
$\partial_\pm = \partial_1 \pm i \partial_2$. One can obtain
$r^2 - 1$ coupled nonlinear second order differential equations
for the $r$ variables $W_\psi$ by plugging the expression for 
$A_\pm^a$ into the third equation of (\ref{eq20}). It appears
at this point that these second order equations form an
over-determined system and, thus are inconsistent. However,
this is not the case as can be verified in a particular gauge.
In the Coulomb gauge,
\be
A_i^a = \pm \epsilon_{ij} \partial_j \zeta^a,
\label{eq22}
\ee
\noindent the first equation of (\ref{eq20}) can be solved as,
\be
W_\psi = exp \ ( \zeta^a \Gamma^a ) \ f ( x_\pm ), \ \
x_\pm = x_1 \pm i x_2 ,
\label{eq23}
\ee
\noindent where $f(x_\pm)$ is a $r$-component column vector with
each component being an arbitrary (anti)holomorphic function.
Using Eqs. (\ref{eq22} ) and (\ref{eq23}) into
the third Eq. of (\ref{eq20}), one is left with $r^2-1$ second order
differential equations in terms of the $r^2-1$ variables $\zeta^a$. 
These equations are highly nontrivial for making any further progress.

Following the discussions of Ref. \cite{klee}, we now show in a particular
gauge that Eqs. (\ref{eq20}) and (\ref{eq21}) reduce to the
corresponding equations of self-dual gauged $CP^1$ model
considered in Ref. \cite{gg}. In particular,
we choose $W_\psi^T = (0, 0, \dots, \chi)$, where $\chi$ is a
single-valued regular complex scalar field\footnote{ The Coulomb gauge
condition (\ref{eq22}) is not implemented for the discussion of this
section.}. With this choice of
$W_\psi$, $W_\psi^\dagger \Gamma^a W_\psi$ and $W_\psi^\dagger
\Gamma^a \partial_\pm W_\psi$ are non-zero for the diagonal
generator $T^D = diag [1, \dots, 1, -(r-1)]/\sqrt{2 r (r-1)}$ only.
Now its trivial to see that Eqs. (\ref{eq20}) and (\ref{eq21}) indeed
reduce to the corresponding equations of self-dual gauged $CP^1$
model \cite{gg}, except for the constant factor $\beta$ which is
$\frac{1}{\kappa}$ for the $CP^1$ case. In this gauge, the $U(1)$ charge
is determined in terms of the magnetic flux $\Phi^D= \int d^2 x
F_{12}^D$ as $\tilde{Q} = - \kappa \Phi^D$. In general, the magnetic
flux $\Phi^D$ is not quantized \cite{gg} in terms of the degree of
the map $\Pi_2(CP^N)=Z$. Thus, the self-dual solutions
are of four different types: (i) topological, (ii) nontopological,
(iii) Q-lumps and (iv) Q-balls.
 
To conclude, we have considered a $CP^N$ model with its subgroup
$SU(r) \ ( 1 < r < N+1 )$ completely gauged, where the gauge field
dynamics is solely governed by a nonabelian CS term and the
scalar field self-interaction is given by a $SU(r)$ invariant potential.
We obtained self-dual equations of this gauged $CP^N$ model, where the
energy is bounded from below by a linear combination of the topological
charge and a global $U(1)$ charge present in the theory. The self-dual
soliton solutions of this model are expected to be of four different types,
namely, topological, nontopological, Q-balls and Q-lumps. 
Similar self-dual gauged $CP^N$ models, with a different global $U(1)$
symmetry from the one presented in this paper, are also considered
in Ref. \cite{cp2}. It would be nice if the self-dual equations of these
models can be analyzed in a general frame-work to get better insight into 
the nature of the soliton solutions.


\begin{references}

\bibitem{sig} A. A. Belavin and A. M. Polyakov, JETP Lett. 22
(1975) 245; V. L. Golo and A. M. Perelomov, Phys. Lett. B 79
(1978) 112; R. Rajaraman, Solitons and Instantons (North Holland,
Amsterdam, 1982); W. J. Zakrzewski, Low Dimensional Sigma Models
(Adam Hilger, Bristol, 1989).

\bibitem{nardeli} G. Nardelli, Phys. Rev. Lett. 73 (1994) 2524;
Phys. Rev. D 52 (1995) 5944; Y. M. Cho and K. Kimm, Phys. Rev. D
52 (1996) 7325 ; P. K. Ghosh, MRI/PHY/01/96, hep-th/9601009, To
appear in Phys. Lett B.

\bibitem{may} B. M. A. G. Piette, D. H. Tchrakian and
W. J. Zakrzewski, Phys. Lett. B 339 (1994) 95 and references
therein.

\bibitem{dur} B. J. Schroers, Phys. Lett. B 356 (1995) 291;
J. Gladikowski, B. M. A. G. Piette and B. J. Schroers,
Phys. Rev. D 53 (1996) 844.

\bibitem{gg} P. K. Ghosh and S. K. Ghosh, Phys. Lett. B 366 (1996)
199; K. Kimm, K. Lee and T. Lee, Phys. Rev. D 53 (1996) 4436.

\bibitem{itf} B. J. Schroers, ITF-96-05, hep-th/9603101.

\bibitem{klee} K. Lee, Phys. Rev. Lett. 66 (1991) 553;
Phys. Lett. B 255 (1991) 384.

\bibitem{bogo} E. B. Bogomol'nyi, Yad. Fiz. 24 (1976) 861
[Sov. J. Nucl. Phys. 24 (1976) 449].

\bibitem{rl} R. A. Leese, Nucl. Phys. B 344 (1990) 33;
Nucl. Phys. B 366 (1991) 283. 

\bibitem{sc} S. Coleman, Nucl. Phys. B 262 (1985) 263.

\bibitem{cp2} K. Kimm, K. Lee and T. Lee, CU-TP-742, SNUTP-96-24,
hep-th/9603128.

\end{references}
\end{document}